\documentclass[twocolumn]{article}
\pdfoutput=1

\usepackage[english]{babel}
\usepackage[utf8]{inputenc}

\usepackage{graphicx}
\usepackage{booktabs}
\usepackage{csquotes}

\usepackage[hyphens]{url}
\usepackage[bookmarks=false]{hyperref}
\hypersetup{
            colorlinks=true,
            urlcolor=[rgb]{0.0,0.1,0.75},
            citecolor=[rgb]{0.0,0.1,0.75},
            linkcolor=[rgb]{0.0,0.1,0.75}
}
\usepackage[
backend=bibtex,
natbib=true,
style=authoryear,
sorting=nty,
citestyle=authoryear,
dashed=true,
urldate=long,
firstinits=false
]{biblatex}
 
\usepackage[para,online,flushleft]{threeparttable}

\makeatletter
\renewcommand\section{\@startsection{section}{1}{\z@}%
                                  {-3.5ex \@plus -1ex \@minus -.2ex}%
                                  {2.3ex \@plus.2ex}%
                                  {\normalfont\large\bfseries}}
\renewcommand\subsection{\@startsection{subsection}{1}{\z@}%
                                  {-3.5ex \@plus -1ex \@minus -.2ex}%
                                  {2.3ex \@plus.2ex}%
                                  {\normalfont\normalfont\bfseries}}
\makeatother

\begin{document}

\date{\textit{Draft}, \today}

\title{New Technology Assessment in Entrepreneurial Financing - \\ Can Crowdfunding Predict Venture Capital Investments?}

\author{Jermain Kaminski\footnote{RWTH Aachen, Technology Entrepreneurship Group; MIT Media Lab, Macro Connections Group.}, Christian Hopp\footnote{RWTH Aachen, Technology Entrepreneurship Group.}, Tereza Tykvova\footnote{Hohenheim University, Chair of Corporate Finance; Centre for European Economic Research (ZEW).}}

\maketitle
\hspace{-2.45ex}\textbf{Abstract}\\
Recent years have seen an upsurge of novel sources of new venture financing through crowdfunding (CF). We draw on 54,943 successfully crowdfunded projects and 3,313 venture capital (VC) investments throughout the period 04/2012-06/2015 to investigate, on the aggregate level, how crowdfunding is related to a more traditional source of entrepreneurial finance, venture capital. Granger causality tests support the view that VC investments follow crowdfunding investments. Cointegration tests also suggest a long-run relationship between crowdfunding and VC investments, while impulse response functions (IRF) indicate a positive effect running from CF to VC within two to six months. Crowdfunding seems to help VC investors in assessing future trends rather than crowding them out of the market.\\
\\
\textbf{Keywords}\\
Crowdfunding, venture capital, Granger causality, crowding out.\\
\\
\textbf{Classification}\\
G30, G24, O3\\
\\
\section{Introduction}
\vspace{5mm} 
"[T]his year, we've been slower to invest partially because in our analysis, there are years where there are lots of new ideas and big swings that are going for new industries. I feel like last year and maybe the year before were better years for big new ideas. This year, we haven't seen as many."\footnote{\url{http://techcrunch.com/2015/09/28/early-stage-investors-apply-the-brakes}; accessed June 1st, 2016.}
\begin{flushright}
\vspace{2mm} 
\textit{Aileen Lee, Cowboy Ventures}
\end{flushright}
\vspace{5mm} 

Entrepreneurship is “always a voyage of exploration into the unknown, an attempt to discover new ways of doing things better than they have been done before” \citet[p. 101]{hayek_1948}. The need for experimentation is deeply rooted in entrepreneurship. An environment which facilitates experimentation and tolerates failures is the mainstay of a prospering economic ecosystem that generates innovations \citep{kerr_nanda_rhodes_kropf_2014}.

Within such an environment, many new business ideas compete, but only a few single ventures will actually survive and bring their products to market. As future outcomes are distant und unpredictable, entrepreneurs and their financiers are plagued with uncertainty about whether or not a particular product will sell or a business model will turn out to be successful. Different types of financiers deal with these uncertainties in different ways. In crowdfunding, many small investors place bets on the potential of a business idea. To receive venture capital (VC) funding business ideas undergo a profound selection process through a specialized investor. The aim of this paper is to investigate  how crowdfunding relates to VC. More specifically, we seek to answer the question whether crowdfunding and VC are rather complements or substitutes. 

Crowdfunding is a relatively new gateway for entrepreneurs to access capital for creative and novel ideas
\citep{belleflamme_lambert_schwienbacher_2014, cumming_johan_2016, mollick_2014, schwienbacher_larralde_2012}. It allows individuals and creative users to start experiments with new products and services \citep{vonhippel_1986} within a wide range of capital supply. The amount can vary from a few dollars for a pottery project to \$20 million for a new smart watch. Average individual contributions are rather low\footnote{\citet{mollick_2014} reports the average goal of \$5,064 for all 2,317 funded projects on Kickstarter in the period from its inception in 2009 to July, 2012; the average contribution amounts to \$20,563 in our sample.}  and entrepreneurial ventures typically use crowdfunding in the very early, exploratory phase. The money is raised via online platforms, such as Kickstarter, which provide entrepreneurs with a direct interaction with potential end users. Through their feedback, end users provide evaluations of the business idea and may prove useful in reshaping and improving the product. The entrepreneurs not only attract potential customers and early-stage financiers on the crowdfunding platform, but also obtain higher visibility, for example through press coverage. Crowdfunding may help to aggregate market expectations about the likelihood of future events, trends and technologies. \citet{mollick_kuppuswamy_2014} suggest that crowdfunding appears to support traditional entrepreneurship.
In fact, if the crowd provides expert-like assessment of future technology trends, than a positive long-run relationship between crowdfunding and VC investments may exist, in which VC investments follow crowdfunding investments at a time lag. 

Crowdfunding processes may validate the feasibility and viability of new technologies. Crowdfunding not only reflects an investment into future products, i.e., needs that may not be satisfied in the current market, but also provides VC investors with a crowd-driven proof of concept for certain technologies and markets. Just like bees may indicate the direction to a distant flower patch yielding nectar and pollen through a ‘waggle dance’ in their hive \citep{vonfrisch_1967}, crowdfunding investors may fulfil a complementary role to VC investors, by providing them with information towards new products and product categories. Crowdfunding platforms may thus support products and ideas, which would otherwise not be pursued. The “wisdom of the crowd” may not only lead to more projects that search for VC financing, but also improve the quality of projects that are in the market. This is important as deal flow has been shown as a large concern for VC investors \citep{hochberg_ljungquist_lu_2010}, in particular in ho periods when “too much money chases too few deals” \citep{gompers_lerner_2000}. Alternatively, we consider the opposite view, namely that crowdfunding and VC investors compete for deals in some segments of the market, where the availability of crowdfunding increases the supply of funds that compete for a limited number of projects. If this is the case, crowdfunding and VC investments will be substitutes and crowdfunding would crowd out VC investments.

Our research questions ask whether there is a long-run relationship between crowdfunding and VC investments on the aggregate and the industry levels, and whether crowdfunding and VC investments are linked, such that an increase in crowdfunding causes an increase (or decrease) in VC investments. To answer these questions, we use a dataset covering 54,943 crowdfunding investments that have successfully reached their funding goal and 3,313 VC investments in the US. Our analysis starts with the adoption of JOBS Act in April 2012, which fueled crowdfunding investments, and ends in June 2015. We test whether one time series Granger causes the other, and we employ a cointegration test to investigate whether there is a long-run relationship between crowdfunding and VC investments. The results suggest that crowdfunding Granger causes VC investments. We also find a long-term relationship between crowdfunding and VC investments; the time series are cointegrated. Successful crowdfunding campaigns lead to a subsequent increase in VC investments. This holds in aggregate and especially for hardware and consumer electronics.

We perform a separate analysis for ‘small’ VC investments below \$500,000 (which are comparable to crowdfunding investments in their size) to investigate whether crowdfunding may potentially crowd out VC investments within this segment. We do not find evidence in support of a crowding-out effect. 
Our work helps to clarify the role that crowdfunding plays in financing entrepreneurial ventures. It contributes to several strands of the literature. We add to the emerging literature on crowdfunding
\citep[e.g.][]{agrawal_catalini_goldfarb_2015,ahlers_cumming_guenther_schweizer_2015,lehner_grabmann_ennsgraber_2015,cumming_johan_2016,mollick_2013,mollick_kuppuswamy_2014,mollick_nanda_2015}. To the best of our knowledge, no empirical analysis exists that focuses on the possible connection between crowdfunding and VC investments, though recent work addresses the link between business angels and VC investors \citep{berkovitch_grinstein_israel_2015,chemmanur_chen_2014,hellmann_schure_vo_2015,hellmann_thiele_2015}.

Our study extends previous analyses that suggest that crowdfunding is able to lead to and support traditional entrepreneurship\citep{mollick_kuppuswamy_2014}. The (visible and oftentimes large scale) involvement of the online brand community may present a strong signal of a project viability and market validity, which, in turn, may influence the perception of other investors that have not invested yet. Our study seems to support the view that crowds are not ‘mad’ but rather ‘wise’ \citep{mollick_nanda_2015}.
As such, our work is linked to theoretical work focusing on equilibrium models that endogenize the size and structure of the seed/angel stage market to conceptually model the size of the VC market \citep[e.g.][]{hellmann_thiele_2015}. VC funded startups might jump on a bandwagon of new trends discovered on crowdfunding platforms and commercialize these ideas with more financial resources that enable to scale faster. Hence, the existence of crowdfunding encourages entrepreneurs to engage in more experimentation, which subsequently broadens the market for VC investments. 

It is not only the quantity, but the early test for viability may also improve the quality of projects that compete in the marketplace. Crowdfunded entrepreneurs might generate spillovers and spur the birth and development of further companies in the same and other technology fields as successful crowdfunding campaigns can create interest in new projects and co-creations.

\section{Related Literature and Research Question}

\subsection{Entrepreneurship as Experimentation}

In March 2014, when Facebook bought Oculus Rift – a company which raised money via Kickstarter –, Marc Zuckerberg stated: “One day, we believe this kind of immersive, augmented reality will become a part of daily life for billions of people. Virtual reality was once the dream of science fiction. But the internet was also once a dream, and so were computers and smartphones.”\footnote{\url{https://www.facebook.com/zuck/posts/10101319050523971}, accessed June 1st, 2016.}

At the origin of such radically new products or services, there are creative and innovative entrepreneurs, who, instead of being guided by what is in the present, arrange their affairs on the grounds of their visions for the future. An entrepreneur imagines a novel product or service that would excite customers and generate large market demand. “What he ‘sees’ is that, by assembling available resources in an innovative, hitherto undreamt of, fashion, and thus perhaps converting them into new, hitherto undreamt of products, he may be able (in the future) to sell output at prices which exceed the cost of that output to himself”\citep[p. 150]{kirzner_2009}.

Yet, for novel firms and unproven technologies, tremendous uncertainty exists about whether the revenue model will work and whether the technology will be feasible. The probability that such endeavors will succeed is low, extremely skewed, and impossible to predict (Kerr et al., 2014a). In a Canadian sample, \citep{aastebro_2003} finds that only 7-9\% of entrepreneurs succeed in turning their inventions into marketable products. Of those few that do, 60\% earn negative returns, while 8\% reach returns higher than 1,400\%. Related to this, \citet{cochrane_2005}, investigating the returns of VC investors, finds that they are, by and large, driven by a handful of very successful investments.

Experimentation with various business ideas is necessary to find these very successful investments and to separate the good from the bad opportunities. Recent literature \citep{kerr_nanda_rhodes_kropf_2014,manso_2011,nanda_rhodes_kropf_2013} suggests that important factors (beyond financial capital) that drive entrepreneurial activity are a mindset of experimentation and a willingness to fail. To create radically novel products and services, entrepreneurs – who act in good faith on imagined new venture ideas – must have the chance to experiment with their ideas and not to be punished when they fail \citep{manso_2011}. Recent work strongly points towards the benefits of experimenting with diverse business ideas, which results in more ideas coming to fruition, and has a longer-lasting effect on subsequent novel activity \citep{kornish_ulrich_2011,ostergaard_timmermans_kristinsson_2011} and technological progress \citep{kerr_nanda_rhodes_kropf_2014}. 

\subsection{Entrepreneurship as Experimentation}

The availability of financial capital is recognized as having a significant impact on whether entrepreneurs are able to progress in the gestation stage of new venture creation, or whether they have to abandon their business idea \citep{holtz_eakin_joulfaian_rosen_1994, reynolds_2011, vangelderen_thurik_bosma_2006}. For VC investors, it is difficult to assess which technology or venture will succeed or which industry offers a high potential \citep{ewens_rhodeskropf_2015}. In fact, strong variations exist in how skilled VCs are at assessing potential investments \citep{agrawal_catalini_goldfarb_2015}). In consequence, substantial residual uncertainty about venture viability exists. \citet{kerr_lerner_schoar_2014} argue that the highly skewed returns of VC investors are evidence in favor of high uncertainty and large prediction errors, such that even the more prominent VC investors cannot distinguish the next Steve Jobs from the next Adam Osborne. Not surprisingly then, \citet{kerr_lerner_schoar_2014} report low correlations between VC investors’ assessments and levels of future success. VC investors passing on what later turns out to be extremely successful endeavors are not an exception. \citet{kerr_lerner_schoar_2014} describe the hesitations of VC investors to fund Airbnb because they could simply not understand (or imagine) that the business model could actually work. 

When VC investors select in which companies and industries to invest, they consider different criteria. \citet{kaplan_stroemberg_2004} distinguish between three categories: internal factors (such as management quality), external factors (such as market size and customer adoption) and difficulty of execution and implementation. While internal factors are subject to asymmetric information between the entrepreneur and the VC investor, external factors are subject to uncertainty, but information is not necessarily asymmetrically distributed because entrepreneurs may face the uncertainty as well. VC investors often reward stock options to founders that secure threshold number of customers who have purchased the product and reported positive feedback \citep{kaplan_stroemberg_2004}. This contingency supports the view that external factors, such as the expected market size and customer adoption, are important for VC investors. This is understandable, because they directly affect project revenues. 

\subsection{Technology Assessment and Crowdfunding}

Crowdfunding offers multiple evaluations by thousands of potential backers, who, as end users of the product, might be well-positioned to judge the project’s viability. Through their investments, individual backers express their interest in using the product and their belief in its successful development. Products are therefore not developed based on the perceptions and wishes of imaginary customers. Instead, they are developed if and only if real customers buy into their vision today. 
Prior research has found that aggregated group decisions tend to be more accurate than decisions by individuals in which only a single decision maker relies on his sole assessment \citep{budescu_chen_2014,larrick_mannes_soll_2011}. Not surprisingly, bandwagon effects have been reported \citep{mollick_2013}, where additional investors come on board when momentum is building up. A wider range of expertise makes evaluation more accurate, reduces information frictions, and, thus, provides greater efficiency in funding decisions. The crowd has been reported to be surprisingly rational in their decision making, despite the potential for herding and madness \citep{mollick_nanda_2015}. Recent empirical evidence supports the view that the crowd makes accurate, expert-like assessments \citep{mollick_nanda_2015}, and relies on signals \citep{ahlers_cumming_guenther_schweizer_2015,mollick_2013}. \citet{mollick_kuppuswamy_2014} demonstrate a large success probability among projects funded via Kickstarter. Over 90\% of funded projects remained ongoing ventures 1-4 years after their campaign. 

By offering a direct market test, crowdfunding makes product demand more calculable. Not surprisingly, crowdfunding has been influential in areas such as gaming, internet-related technologies, wearable computing, and three-dimensional printing \citep{mollick_nanda_2015}, which directly address end users. The visible and large scale involvement of the online brand community may present a strong signal of a project’s viability and market validity, which in turn may influence the perception of other investors that have not invested yet \citep{agrawal_catalini_goldfarb_2015,lehner_grabmann_ennsgraber_2015,cumming_johan_2016}. Also, successful entrepreneurs generate spillovers and spur the evolution of new technology fields, and this initiates the emergence and growth of further companies. 

\subsection{Crowdfunding and Venture Capital Financing}

While information generation for the individual investor is expensive and cumbersome to obtain, observing several experiments and their reception by the market may be valuable and may help to update prior beliefs. While VC investors may be uncertain about the diffusion of new technologies, they may observe crowdfunding behavior and process information to make more informed investment choices. Successful experiments shift the probability predictions of uncertain technologies and may make a continuation or funding decision more likely. When VC investors observe that a certain new technology received widespread community support, they may interpret it as an indicator of the true venture viability and technological feasibility in this field. According to \citep[p. 131][]{zider_1998}, “One myth is that venture capitalists invest in good people and good ideas. The reality is that they invest in good industries.” Crowdfunding campaigns can create interest in new projects and co-creations. Take for example the market for smart locks. While Lockitron raised some \$2 million directly from their customers, the Yves Behar start-up August raised the same amount from Cowboy Ventures (and syndicate partners) and entered the market a few months later. Both companies represent bets on a potential standard that might emerge and whether or not consumers will gravitate to one or another of the very different designs of the products. The VC investors placed their bet after witnessing the early success of the first mover, and scaled their venture faster.\footnote{\url{http://www.hackthings.com/who-will-win-the-smart-lock-race-august-or-lockitron/}, accessed June 1st, 2016.}

Above and beyond the crowd’s assessment of emerging technologies and trends crowdfunding may also increase the number of entrepreneurial projects that will be pursued and thus increase the deal flow for VC investors. Consequently, crowdfunding may encourage entrepreneurs to pursue their experiments and simultaneously provide information on future markets trends (and potential market sizes). 

Alternatively, crowdfunding could also compete with VC investors for funding and in supporting promising business ideas. As backers of projects know their needs and may also possess knowledgeable skills in the technical domain in which the project operates, they may be able to assist the entrepreneur in product development by creating an ecosystem that facilitates access to resources between various stakeholders \citep{belleflamme_lambert_schwienbacher_2014,frydrych_bock_kinder_koeck_2014}. Crowdfunding may therefore provide a viable instrument to finance entrepreneurial experiments. In comparison with ‘real-life’ experimentation, in which the costs of failure and stigma might be high \citep{landier_2005}, the crowd may provide a low-cost alternative for experimentation \citep{mollick_2014}. Similarly, the crowd provides financial capital up-front and not conditional on reaching various milestones as it is common in other types of entrepreneurial financing \citep{tian_2011,tian_wang_2014}. In fact, instead of relying on VC investors to provide value-adding advice, entrepreneurs might simply substitute formal capital with informal capital through the crowd. This might be relevant especially for smaller and earlier investments.

\section{Data}

We draw our sample from the crowdfunding platform Kickstarter and from the VC database CrunchBase. As we explain below, we believe that these two data sources are representative of crowdfunding and VC investment activity in US-based companies within the selected time window. We describe the VC segments, subsegments and categories in Table~\ref{tab:descriptionofvariables}.  

\begin{table*}[t]
\centering
\caption{Investment segments, subsegments and categories.}
\label{tab:descriptionofvariables}

\begin{tabular*}{\textwidth}{@{}l@{\extracolsep{\fill}{\hskip 0.1in}}l@{\extracolsep{\fill}}}
\toprule
{Variable} & {\it Description}\\ 
 \midrule
  \textit{Investment Segments} &    \vspace{4pt}\\
  cf\_sum & Volume (\$) of Kickstarter crowdfunding investments.\\
  vc\_sum & Volume (\$) of CrunchBase venture capital investments (angel, seed, venture).\\
  vc\_angel\_seed & Volume (\$) of venture capital investments attributable to angel investor \\
  &  and seed stage financing, denoted as “Angel” and “Seed” at CrunchBase. \\
  vc\_early\_growth & Volume (\$) of venture capital investments attributable to venture\\
  & (early stage and growth) financing, denoted as “Venture” at CrunchBase. \\
  \textit{Investment Categories} &  \vspace{4pt}\\
  cf\_hardware, vc\_hardware & Volume (\$) of investments in the category hardware and consumer\\
  & electronics (hardware).\\
  cf\_media, vc\_media & Volume (\$) of investments in the category media, arts, entertainment (media).\\
  cf\_fashion, vc\_fashion & Volume (\$) of investments in the category fashion, wellness, personal care \\
  & (fashion).\\
\bottomrule
\end{tabular*}
\end{table*}

Kickstarter is a global crowdfunding platform, which has the mission to “help bring creative projects to life”.\footnote{\url{https://www.kickstarter.com/charter/}, accessed June 1st, 2016.}  Kickstarter serves as an intermediary between potential funders and creators of projects, and does not claim any ownership over the projects. Individual funders (who are called backers) can contribute small amounts of money starting from a few dollars. For their contributions they are offered rewards, which vary depending on the amount the backer contributes; they can include things like cards, t-shirts, cups, or the possibility to spend a day with the project creator or to be one of the first that obtain the new product. Kickstarter operates on an All-or-Nothing basis, meaning that only when funding goals are reached the project creator will receive the pledged funds \citep{cumming_leboeuf_schwienbacher_2014}. In addition, potential and actual project backers may participate in online discussions and exchange their opinions about the project with the creator and among each other. Creators of projects apply for funding in fifteen different areas ranging from culture to technology. When they post a project, they choose a deadline and a minimum funding goal. If the goal is not reached by the deadline, the project ends without any funds collected. We gather information on all projects based in the US that successfully reached their funding goal within the period from 04/2012 to 06/2015 on the Kickstarter platform. This was possible as the webpages of all projects are archived and accessible to the public. Our crowdfunding segment comprises 54,943 projects and a total funding volume of \$1.13 billion (see Table~\ref{tab:descriptivesprojectlevel}) pledged by a total of 14,450,179 backers. A median project raised \$5,206. The average pledge amount of \$20,563 is attributable to several major projects that raised more than \$50,000.

We employ Kickstarter for three main reasons. Firstly, Kickstarter is the world’s largest online crowdfunding platform, having raised more than \$2 billion from its inception on April 28, 2009 to October 31, 2015. Secondly, Kickstarter has been described as being representative of projects seeking funding from the crowd \citep{mollick_nanda_2015,mollick_2014}. Thirdly, Kickstarter campaigns are associated with a strong backer network and active online communication between (potential) end users. 

We perform analyses on the aggregate level (our aggregate crowdfunding measure is based on all 15 categories available on Kickstarter) and, in addition, we focus separately on the following investment categories: hardware and consumer electronics (which we label hardware), media, arts and entertainment (media), and fashion, wellness and personal care (fashion).\footnote{\citet{mollick_2014} defines seven basic themes based on the 15 Kickstarter categories: namely (i) art and craft, (ii) design, (iii) music, dance and theater, (iv) fashion, (v) film, video, and games, (vi) publishing, photography, comics, and journalism, and (vii) technology as the main categories. He does not mention the category of food.}  We are able to assign 94.3\% of all projects to one of these three categories.\footnote{The only Kickstarter category that we do not assign to one of the investment categories is food because it does not fit to any of our three investment categories. We also cannot include it as a separate category because the time series yields several zero entries, which would make the empirical analysis too fragile.}

Our second data source, CrunchBase, is a database which comprises data on VC investments, VC-backed companies, and VC investors in the US \citep{cumming_walz_werth_2016}. Every registered member can make submissions to the database; however, all changes are subject to review by a moderator before being accepted. Analogously to the crowdfunding sample, we extract data for investments in US-based companies during the period spanning from 04/2012 to 06/2015. The VC segment consists of 3,313 VC investments totaling \$19.1 billion (see Table~\ref{tab:descriptivesprojectlevel}). We retrieved the sample on August 17, 2015 for CrunchBase subsegments ‘angel’, ‘seed’, and ‘venture’. Due to the very low number of observations in the angel subsegment, we merged the observations from the first two subsegments into one: angel\_seed. We label the second subsegment early\_growth, as it includes both these stages of VC financing. We trimmed the sample to comprise only the 99\% percentile to avoid size distortions by extreme outliers. 
We are confident that CrunchBase is representative of the VC investment activity in US-based companies. \citet{block_sandner_2009}, who focus on Internet start-ups between 2007 and 2009, report that CrunchBase quarterly time series correlate highly with the National Venture Capital Association (NVCA) quarterly time series (Pearson correlation coefficient of 0.67) and cover almost 97\% of the volume reported by the NVCA.\footnote{We cannot use data from the NVCA for our purposes, because they do not provide monthly time series for investment categories.}  

Within the CrunchBase we find a vast number of investments that do not correspond to similar categories in Kickstarter. We eliminate categories such as biotechnology, software, e-commerce and enterprise software development, clean technology and manufacturing, health care and life science, advertising, and semiconductors from the CrunchBase sample. The remaining categories represent a map with the three investment categories we defined for Kickstarter projects.

To assess the coverage within investment categories, we make a comparison between CrunchBase and other data sources. In the media category, for example, the volume of investments reported in CrunchBase amounts to \$10.8 billion (see Table~\ref{tab:descriptivesprojectlevel}), while the volume of investments reported by NVCA during 2012-2014 in the category media and entertainment amounts to \$11.0 billion \citep{nvca_2015}. Compared to other databases commonly used in VC research, such as Thomson or Venture Source, CrunchBase seems to have a better coverage of smaller investments. For example, within our sample period we find 670 investments in the hardware category in CrunchBase, with a median of \$1.7 million and a 25 percentile of \$0.3 million. In Venture Source, we find only 310 investments in the category electronics and hardware, with a median of \$4.5 million and a 25 percentile of \$1.3 million. In the media category, our sample has a median of \$1.5 million and a 25 percentile of \$0.3 million, while the mean and 25 percentile investments in the categories (i) media and content and (ii) entertainment reported in Venture Source reach \$3.0 million and \$1.1 million respectively. We therefore believe CrunchBase to be representative of the VC investment universe and to have a wider coverage than competing databases. Similarly, \citet{alexy_block_sandner_terwal_2012} attest to the wide spectrum covered in CrunchBase, ranging from very small to very large companies, such as Google and Facebook.

\begin{table*}[t]
\centering
\caption{Descriptive statistics at project level.}
\label{tab:descriptivesprojectlevel}

\begin{tabular*}{\textwidth}{@{}l@{\extracolsep{\fill}}l@{\extracolsep{\fill}}l@{\extracolsep{\fill}}l@{\extracolsep{\fill}}l@{\extracolsep{\fill}}l@{\extracolsep{\fill}}l@{\extracolsep{\fill}}l@{\extracolsep{\fill}}}
\toprule
{Variable} & {Count} & {Volume(\$)} & {Mean(\$)} & {Std. dev. (\$)} & {25\% (\$)} & {50\% (\$)} & {75\% (\$)}\\ 
 \midrule
  \textit{Investment Segment} &    \vspace{4pt}\\
cf\_sum             & 54,943 & 1,129,797,195  & 20,563     & 157,455    & 2,120     & 5,206     & 12,750     \\
vc\_sum             & 3,313  & 19,080,899,413 & 5,759,402  & 10,485,333 & 300,000   & 1,550,000 & 6,000,000  \\
vc\_angel\_seed     & 1,735  & 1,482,544,313  & 854,492    & 1,043,659  & 100,000   & 500,000   & 1,300,000  \\
vc\_early\_growth   & 1,578  & 17,598,355,100 & 11,152,316 & 13,195,885 & 3,000,000 & 6,450,000 & 14,000,000 \\
  &  \\  
  \textit{Investment Category} &    \vspace{4pt}\\

cf\_hardware        & 5,635  & 381,439,823    & 67,691     & 394,824    & 3,865     & 14,374    & 47,942     \\
vc\_hardware        & 670    & 4,041,163,781  & 6,031,588  & 10,258,284 & 300,000   & 1,737,500 & 6,500,000  \\
cf\_media           & 44,125 & 653,974,351    & 14,821     & 101,534    & 2,025     & 4,710     & 10,460     \\
vc\_media           & 1,960  & 10,762,930,264 & 5,491,291  & 10,218,517 & 323,750   & 1,500,000 & 5,500,000  \\
cf\_fashion         & 2,062  & 41,994,805     & 20,366     & 53,100     & 2,494     & 6,679     & 19,060     \\
vc\_fashion         & 503    & 3,646,393,650  & 7,249,292  & 12,285,636 & 500,000   & 2,000,000 & 9,000,000 \\
\bottomrule
\end{tabular*}
\end{table*}

Our analysis considers crowdfunding and VC investments in a timeframe starting with the JOBS Act enactment\footnote{The JOBS Act was signed by Barack Obama on April 5, 2012, and became public law immediately. \url{http://thomas.loc.gov/cgi-bin/bdquery/z?d112:HR03606:@@@X}, accessed June 1st, 2016.} in April 2012. Until the passage of the JOBS Act, the use of internet funding portals in private placements was extremely limited by law. The JOBS Act introduced several important changes in laws and regulations and crowdfunding volumes made a jump after the JOBS Act was enacted. According to our data, it was only in 2012 that the volume of investments through Kickstarter first exceeded \$10 million. 

\section{Methodology}

We perform our analysis in four steps. We begin with correlations between crowdfunding and VC investment time series (for investment segments, subsegments, and categories). Second, we test for the order of integration of all our time series, which, subsequently, affects our choice of methods. We employ the augmented Dickey–Fuller (ADF), Phillips–Perron (PP) and non-parametric Kwiatkowski–Phillips–Schmidt–Shin (KPSS) tests \citep{dickey_fuller_1979,dickey_fuller_1981,kwiatkowski_phillips_schmidt_shin_1992,phillips_perron_1988}. Several authors argue that joint testing of both nulls can strengthen inferences made about the stationarity or non-stationarity of time series \citep[e.g.][]{maddala_kim_1998}.\footnote{The ADF and PP unit root tests test the null hypothesis that a time series \(Y_{t}\) is \(I(1)\), while the KPPS stationarity test is a complementary test for the null hypothesis that \(Y_{t}\) is \(I(0)\). ADF and PP tests can fail to reject the null hypothesis of a unit root for many economic time series (e.g., \citet{davidson_mackinnon_2004}).}  
Third, we focus on Granger causality between crowdfunding and VC investments. In fact, robust inferences following \citet{granger_1969} are only possible with stationary time series. \citet{granger_newbold_1974} as well as \citet{phillips_1986} point out that Granger causality tests of non-stationary or mixed-stationary time series might suggest a statistically significant relationship where there is none: i.e., a ‘spurious regression’. As some of our time series are not stationary, we resort to the robust procedure for integrated or cointegrated time-series proposed by \citet{toda_yamamoto_1995}, which we describe below. We complement these findings using a visualization through impulse response functions. Impulse response functions trace the response of current and future values of one endogenous variable to a one-unit increase of the exogenous variable (Cholesky decomposition).

Fourth, in order to detect common stochastic trends among the analyzed time series pairs, we perform the Johansen cointegration test \citep{johansen_1988,johansen_juselius_1990}. We test for the order of cointegration among our tested pairs of time series for investment segments, subsegments, and categories. The Johansen cointegration test makes the assumption that two \(I(1)\) time series are cointegrated when their linear combination becomes an \(I(0)\) process \citep{engle_granger_1987}. If two time series share a common trend, there will be Granger causality in one or more directions between them \citep{cuthbertson_hall_taylor_1992}. Unlike Granger causality test, cointegration tests cannot establish the direction of causality, but identify cointegrating VARs, i.e., long-run equilibrium relationships. 

\subsection{Granger Causality}

\citet[][p. 127]{wiener_1956} introduced the notion that “for two simultaneously measured signals, if we can predict the first signal better by using the past information from the second one than by using the information without it, then we call the second signal causal to the first one”. \citet{granger_1969} followed this idea of causality in the context of linear autoregressive models of stochastic processes. A bivariate linear autoregressive model of two stationary time series \(X_{t}\) and \(Y_{t}\)  with zero means can be written as following vector autoregressions \citep{granger_1969}:

\begin{equation}
\label{eq1}
X_{t} = a_{1}+\sum_{j=1}^{m}\beta_{j}X_{t-j}+\sum_{j=1}^{m}\gamma_{j}Y_{t-j}+\epsilon_{1t}
\end{equation}

\begin{equation}
\label{eq2}
Y_{t} = a_{2}+\sum_{j=1}^{m}\delta_{j}X_{t-j}+\sum_{j=1}^{m}\varphi_{j}Y_{t-j}+\epsilon_{2t}
\end{equation}

where \(m\) is the maximum number of lagged observations included (the model order), matrices \(\beta,\gamma,\delta\,\varphi\) contain the coefficients of the model (contributions of each lagged observation to the predicted values of \(X_{t}\) and \(Y_{t}\)), and \(\epsilon_{1t}\) and \(\epsilon_{2t}\) are uncorrelated white-noise series with mean-zero sequence. If the variance of \(\epsilon_{1t}\) (or \(\epsilon_{2t}\)) is reduced by the inclusion of lags of \(Y_{t}\) (or \(X_{t}\)), then \(Y_{t}\) (or \(X_{t}\)) Granger causes \(X_{t}\) (or \(Y_{t}\)).\footnote{\citet{granger_1969} formalizes his theory of causality under the condition of two main axioms: (1) Causes precede their effects in time and (2) a time series \(X_{t}\) causes a time series \(Y_{t}\), if the prediction of the future of the time series \(Y_{t}\) can be improved using the combined past values of \(X_{t}\) and \(Y_{t}\) rather than using information from the past of \(Y_{t}\) alone. Formally, \(Y_{t}\) is (Granger) causing \(X_{t}\) (\(Y\Rightarrow{X}\)), when \(\sigma^{2}(X\mid{U})>\sigma^{2}(X\mid\overline{U-Y})\) where \(\sigma^{2}(X\mid{U})\) is the prediction error in the case where all information from the past is used and \(\sigma^{2}(X\mid\overline{U-Y})\) is the prediction error corresponding to the situation when \(Y_{t}\) was excluded from the information set.} This implies that \(Y_{t}\) is causing \(X_{t}\), provided the coefficients in \(\gamma_{j}\) are not zero. The interdependence can be tested by performing an \(F\)-test of the null hypothesis that \(\gamma_{j}=0\), given assumptions of covariance stationarity on \(X_{t}\) and \(Y_{t}\) \citep{geweke_1982}. Similarly \(X_{t}\) is causing \(Y_{t}\), if the coefficients in \(\delta_{j}\) are \(\neq{0}\). If both of these events occur, there is said to be a ‘feedback’ between \(X_{t}\) and \(Y_{t}\).

Yet Granger causality is highly dependent on the order of integration and the estimation of the VAR when its variables are integrated. \citet{granger_1969} assumes that only stationary time series are involved. As our results indicate ambiguity as to the order of integration and cointegration properties, we follow the procedures suggested in \citet{toda_yamamoto_1995}.

\subsection{Toda- and Yamamoto-based Granger causality}

\citet{toda_yamamoto_1995} propose a simple but robust procedure irrespective of the system’s integration or cointegration properties. This procedure requires the estimation of an augmented VAR which guarantees the asymptotic distribution of the Wald statistic (an asymptotic \(\chi ^{2}\)-distribution), since the testing procedure is robust to the integration and cointegration properties of the process \citep{dolado_luetkepohl_1996}. The \citet{toda_yamamoto_1995} test is a modified Wald (MWald) test for linear restrictions on some parameters of an augmented VAR(\(m+d_{max}\)) in levels, where \(m\) is the selected lag order and \(d_{max}\) is the highest order of integration suspected in the system (in our case at most 1). A bivariate vector autoregressive model without deterministic terms can be formalized as

\begin{eqnarray}\nonumber
\label{eq3}
X_{t}&=& a_{1}+\sum_{j=1}^{m}\beta_{j}X_{t-j}+\sum_{j=m+1}^{\mathit{m+d_{max}}}\beta_{j}X_{t-j}\\
&& +\sum_{j=1}^{m}\gamma_{j}Y_{t-j}+\sum_{j=m+1}^{\mathit{m+d_{max}}}\gamma_{j}Y_{t-j}+\epsilon_{1t}
\end{eqnarray}

\begin{eqnarray}\nonumber
\label{eq4}
Y_{t}&=& a_{2}+\sum_{j=1}^{m}\delta_{j}Y_{t-j}+\sum_{j=m+1}^{\mathit{m+d_{max}}}\delta_{j}Y_{t-j}\\
&& +\sum_{j=1}^{m}\varphi_{j}X_{t-j}+\sum_{j=m+1}^{\mathit{m+d_{max}}}\varphi_{j}X_{t-j}+\epsilon_{2t}
\end{eqnarray}

where \(\beta,\gamma,\delta\,\varphi\) are the parameters of the model, \(\epsilon_{1t}\) and \(\epsilon_{2t}\) are error terms and assumed to be white noise with zero mean, constant variance and no auto-correlation. In the first equation, the null hypothesis of non-causality from \(Y_{t}\) to \(X_{t}\) can be expressed as \(H_{0}=\gamma_{j}=0,\forall i=1,2,.. m\). Thus, causality implies \(\gamma_{j}\neq0\).

To conduct the Toda and Yamamoto Granger causality test, we employ three sequential procedures:

\begin{enumerate}
\item{We determine the maximum order of integration \(d_{max}\)  for each pair of our variables.  For all VAR models, we conclude that \(d_{max}=1\).}
\item{We ensure the fit of our model through an appropriate lag length selection \(m\) , allowing for a drift and trend in each series. We determine the optimal lag order to ascertain that the VAR is well-specified by, first, minimizing one of the common information criteria Akaike information criterion (AIC), Bayesian information criterion (BIC) or Schwartz information criterion (SIC) and Final Prediction Error (FPE), and, second, taking care of the remaining serial correlation in the residuals through a \citet{ljung_box_1978} portmanteau test (asymptotic). Subsequently, we check on the inverse roots of AR characteristics of polynomials for dynamic stability.}
\item{To test for the robustness of our estimator (measurement with the optimum lag length (\(m\)) we test with \(m+1\) and \(m-1\) to check if the causality among different lags is volatile or whether it follows a smooth transition.}
\end{enumerate}

\section{Results}

\begin{figure*}[t]
\includegraphics[width=\textwidth]{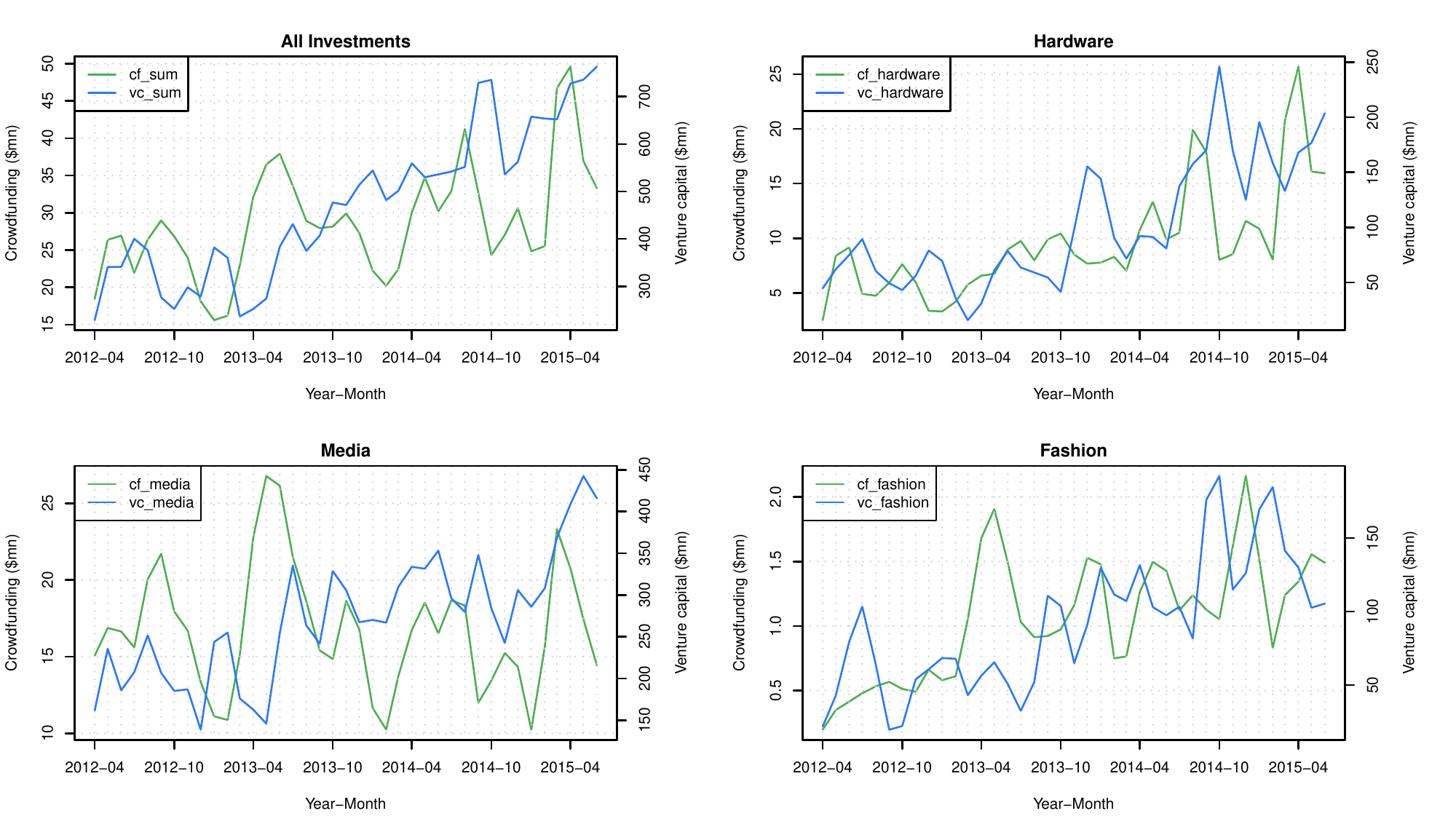}
\caption{Time series of monthly volumes of crowdfunding and venture capital investments.}
\label{fig:timeseries}
\end{figure*}

\subsection{Time Series Correlation}

Figure~\ref{fig:timeseries} shows the development of our time series. The upper left graph depicts the development of the monthly volume of aggregate VC investments and crowdfunding during the sample period. The two time series are positively correlated (Pearson \(\rho\) = 0.431 and the correlation is highly statistically significant \(p\) = 0.006. The remaining graphs show monthly volumes of VC investments and crowdfunding in our three investment categories. The upper right graph visualizes the monthly volumes in the hardware category for 670 VC investments that sum to \$4.0 billion and for 5,635 successful Kickstarter campaigns that raised \$381.4 million (see Table~\ref{tab:descriptionofvariables}). The correlation between the two time series is again positive and highly statistically significant (\(\rho\)=0.553, \(p\)=0.001). The lower left graph depicts monthly investment volumes in the largest investment category, media. There are in sum 1,960 VC investments totaling \$10.8 billion, while 44,125 Kickstarter campaigns raised \$654.0 million. The times series are not correlated. Lastly, the lower right graph depicts the time series for fashion. Over the course of the analysis horizon, we count 503 VC investments totaling \$3.6 billion, while at the same time, 2,062 Kickstarter campaigns raised a total of \$42.0 million. The correlation between the two time series is positive and statistically significant (\(\rho\)=0.357, \(p\)=0.019).

\subsection{Order of Integration}

Considering a 5\% significance level, we find all time series in the aggregate and within the categories to be integrated of first order (as evidenced by the ADF test). The PP and KPSS tests indicate that some time series (cf\_sum, cf\_hardware, cf\_media, vc\_media and vc\_fashion) may be integrated \(I(0)\) when taking a drift and a trend process into account. The time series vc\_sum, vc\_hardware and cf\_fashion are the only time series that are conclusively first difference stationary \(I(1)\). Taken together, our results (not depicted, but available upon request) lack convincing evidence to decisively reject the possibility of no unit root, neither for crowdfunding nor for VC. Our time series are either weakly stationary at level or integrated at first order \(I(1)\).

\subsection{Granger Causality \citep{toda_yamamoto_1995}}

\begin{table*}[t]
\centering
\caption{Toda and Yamamoto (1995) modified Wald (MWald) test for Granger causality.}
\label{tab:granger1}
\begin{tabular*}{\textwidth}{@{}l@{\extracolsep{\fill}}l@{\extracolsep{\fill}}l@{\extracolsep{\fill}}l@{\extracolsep{\fill}}l@{\extracolsep{\fill}}l@{\extracolsep{\fill}}l@{\extracolsep{\fill}}l@{\extracolsep{\fill}}}
\toprule
{Pair} & {Lag} & {\(\chi ^{2}\)} & {\(p\)-value} & {Granger causality}\\
\midrule
\textit{Investment Segments} & \vspace{4pt} \\
cf\_sum$\rightarrow$vc\_sum & 5 & 13.1 & 0.023$^{**}$ & Unidirectional causality \\
vc\_sum$\rightarrow$cf\_sum & 5 & 1.6 & 0.910 & cf\_sum$\rightarrow$vc\_sum \vspace{5pt}\\
cf\_sum$\rightarrow$vc\_angel\_seed & 4 & 15.0 & 0.005 $^{***}$ & Unidirectional causality \\
vc\_angel\_seed$\rightarrow$cf\_sum & 4 & 4.7 & 0.320 & cf\_sum$\rightarrow$vc\_angel\_seed \vspace{5pt}\\
cf\_sum$\rightarrow$vc\_early\_growth & 5 & 13.4 & 0.020  $^{**}$ & Unidirectional causality \\
vc\_early\_growth$\rightarrow$cf\_sum & 5 & 1.4 & 0.920 & cf\_sum$\rightarrow$vc\_early\_growth \vspace{5pt}\\
\textit{Investment Categories} & \vspace{4pt} \\
cf\_hardware$\rightarrow$vc\_hardware & 4 & 11.2 & 0.025$^{**}$ & Unidirectional causality \\
vc\_hardware$\rightarrow$cf\_hardware & 4 & 1.7 & 0.780 & cf\_hardware$\rightarrow$vc\_hardware \vspace{5pt}\\
cf\_media$\rightarrow$vc\_media & 5 & 4.9 & 0.420 & No causality \\
vc\_media$\rightarrow$cf\_media & 5 & 2.5 & 0.770 & & \vspace{5pt}\\
cf\_fashion$\rightarrow$vc\_fashion & 5 & 2.5 & 0.770 & No causality \\
vc\_fashion$\rightarrow$cf\_fashion & 5 & 5.4 & 0.360 & &\vspace{5pt}\\
\bottomrule
\end{tabular*}
\begin{tablenotes}
\item\footnotesize{$^{***}\ p<0.01$, $^{**}\ p<0.05$, $^{*}\ p<0.1$ significance level. \(\chi ^{2}\) and \(p\)-value are MWald test results. Tested with constant and trend at \(maxlag = 6\). Optimal lags are based on VAR estimations (AIC, BIC/SIC, FPE), \citet{ljung_box_1978} portmanteau test (asymptotic) and inverse roots of AR characteristics of polynomials for dynamic stability.}
\end{tablenotes}
\end{table*}

In Table~\ref{tab:granger1} we report the \(\chi ^{2}\)-test statistics and \(p\)-values obtained from the Toda-Yamamoto-based Granger causality MWald tests (against the null hypothesis of no Granger causality). The results indicate that there is a strong Granger causality running from \textit{cf\_sum} to \textit{vc\_sum} (\(\chi ^{2}\)=13.1, \(p\)=0.023). This causality is unidirectional, as we cannot reject the hypothesis of non-causality from \textit{vc\_sum} to \textit{cf\_sum} (\(\chi ^{2}\)=1.6, \(p\)=0.910).\footnote{Additional analyses show that the p-values seem to be a little lower around the optimal lag order of 5. For example, from cf\_sum to vc\_sum, we obtain for lag 4: \(p\)=0.11, for lag 5: \(p\)=0.02, and for lag 6: \(p\)=0.13, while vc\_sum significantly rejects the null hypothesis. This is indicative of a rather robust model, whereas a less robust model would be expected to show less symmetric fluctuations of \(p\)-values for \(m-1\) and \(m+1\).} We also find a strong unidirectional causality in both VC subsegments, i.e. from \textit{cf\_sum} to \textit{vc\_angel\_seed} (\(\chi ^{2}\)=15.0, \(p\)=0.005) and from cf\_sum to vc\_early\_growth (\(\chi ^{2}\)=13.4, \(p\)=0.020). We find a similar result for the investment category hardware as the null hypothesis of non-causality running from \textit{cf\_hardware} to \textit{vc\_hardware} is also rejected (\(\chi ^{2}\)=11.2, \(p\)= 0.025), but the non-causality in the other direction cannot be rejected. Our findings indicate no causality for the media and fashion investment categories. Altogether, the total of crowdfunding investments and in particular crowdfunding in the hardware category, Granger causes respective VC investments.
\begin{figure}[t]
      \includegraphics[width=\columnwidth]{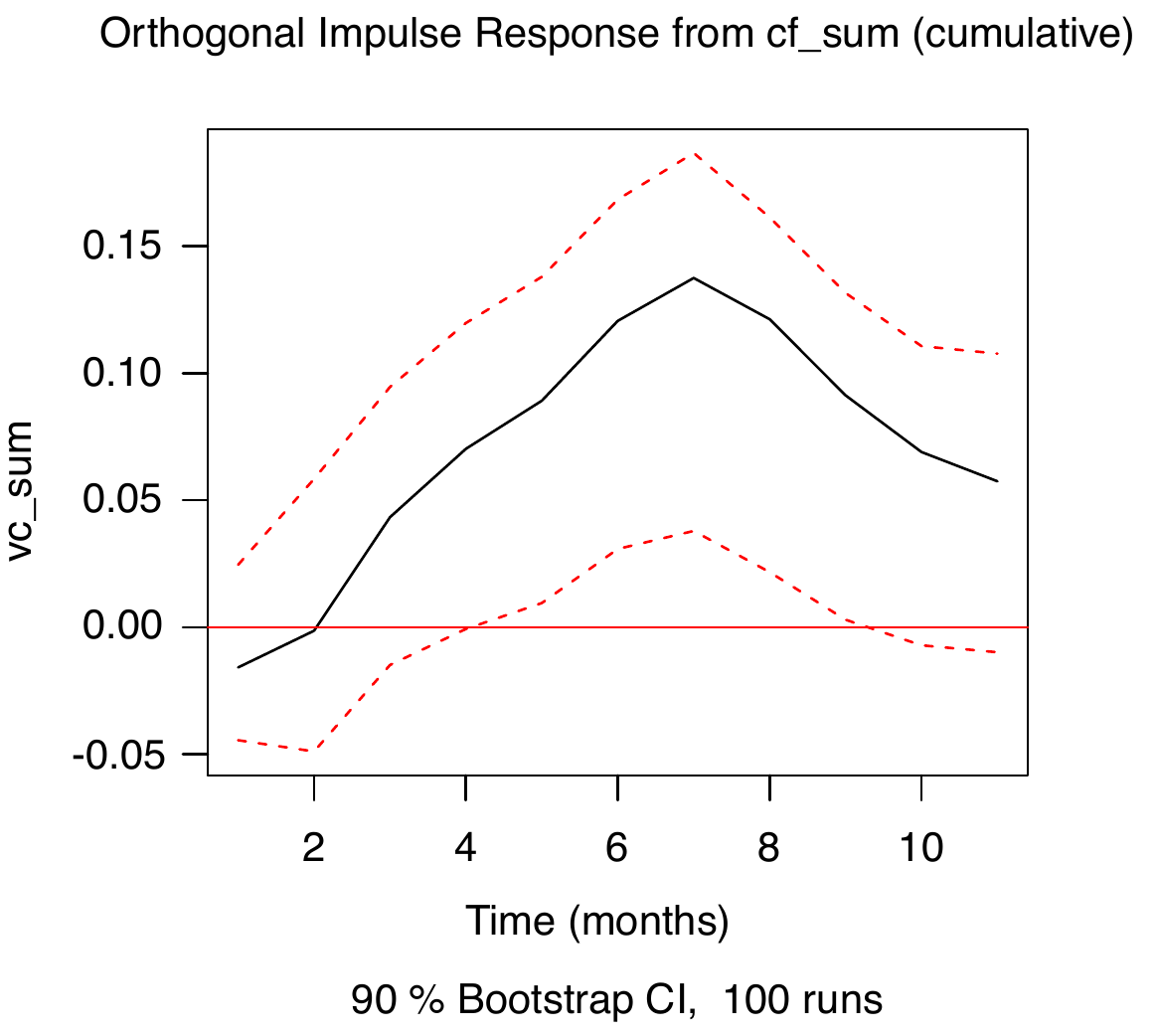}
      \caption{Cumulative impulse response functions of vc\_sum to a 1\% log-increase in cf\_sum. Dashed lines indicate the 90\% bootstrapped confidence interval. Cholesky decomposition for a 2-variable VAR with optimal number of lags chosen with FPE.}
      \label{fig:irf1}
\end{figure}

\begin{figure}[t]
      \includegraphics[width=\columnwidth]{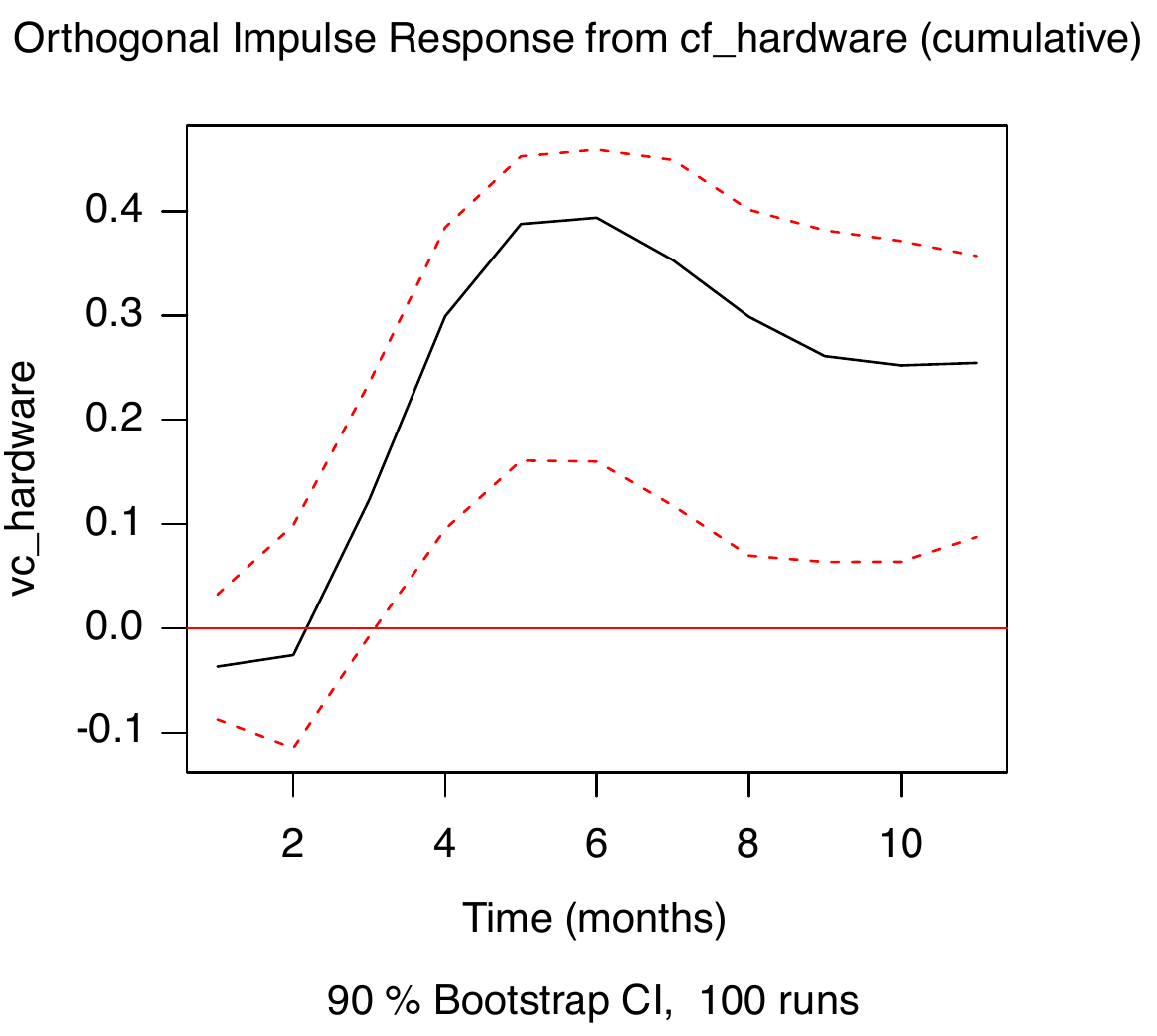}
      \caption{Cumulative impulse response functions of vc\_hardware to a 1\% log-increase in cf\_hardware. Dashed lines indicate the 90\% bootstrapped confidence interval. Cholesky decomposition for a 2-variable VAR with optimal number of lags chosen with FPE.}
      \label{fig:irf2}
\end{figure}
In order to investigate the dynamic effects of changes in crowdfunding on VC investments, we further apply an impulse response analysis (IRF). In Figure~\ref{fig:irf1}, we consider the effects of a 1\% shock of cf\_sum (positive deviation from the steady state) on vc\_sum, and in Figure~\ref{fig:irf2} we look at the effects of a 1\% shock of cf\_hardware on vc\_hardware. For each figure, the black line indicates the point estimate and red dashed lines indicate the two-sided 90\% confidence interval. A 1\% increase in total crowdfunding investments effects a cumulative 4\% increase in total VC investments at lag (month) 4 and a cumulative 10\% increase at lag 6, respectively. There is a positive effect on vc\_sum from lag 2 until lag 7 (with a total increase of 14 basis points). Similarly, vc\_hardware responds positively to a 1\% increase in cf\_hardware. 

\pagebreak
Here, we observe a persistent positive effect on vc\_hardware until lag 6 and a total cumulative increase of 40 basis points. Overall, the impulse response analysis shows a positive effect of crowdfunding investments on VC investments at a lag between 4 and 6. This result resonates well with the outcomes from the Toda-Yamamoto-Granger causality analyses.

\subsection{Cointegration}

Under the assumption that our time series are \(I(1)\) and integrated of the same order, the bivariate Johansen cointegration trace and max eigenvalue test indicate (at a 1\% and 5\% level) that there are cointegration relationships among our analyzed time series pairs (Table~\ref{tab:cointegration}). For cf\_sum and vc\_sum, we can reject the null hypothesis that there is no cointegration vector (\(r\)=0) among the two time series. Similarly, we find at least one cointegration vector for cf\_hardware and vc\_hardware. For the media and fashion investment categories, we find no cointegration vectors.\footnote{These results were confirmed using the bounds testing approach developed by \citet{pesaran_shin_smith_2001}, which is based on the OLS (ordinary least squares) estimation of an ARDL (Autoregressive Distributed Lag) equation. The model in the sense of \citet{pesaran_shin_smith_2001} is not only capable to work with both  \(I(0)\) and  \(I(1)\) variables (as found in the PP and KPSS test) but also more stable towards small sample sizes \citep{pesaran_shin_smith_2001}.}

In sum, the cointegration results support the robustness of the found causality. The Granger causality finds causality where the cointegration test finds cointegration. As we can also reject the null of non-causality in a cointegrated model, we are confident about both results.

\begin{table*}[t]
\centering

\caption{Bivariate Johansen cointegration test.}
\label{tab:cointegration}

\begin{tabular*}{\textwidth}{@{}l@{\extracolsep{\fill}}l@{\extracolsep{\fill}}l@{\extracolsep{\fill}}l@{\extracolsep{\fill}}l@{\extracolsep{\fill}}l@{\extracolsep{\fill}}l@{\extracolsep{\fill}}l@{\extracolsep{\fill}}}
\toprule
{Pair} & {Lag} & {CIV} & {Eigenvalue} & {Trace} & {Max eigenval.} & {Conclusion}\\
\midrule
\textit{Investment Segments} & \vspace{4pt} \\
cf\_sum, vc\_sum & 5 & r=0 & 0.481 & 36.131$^{***}$ & 22.284$^{**}$ & At least one coint. vector. \\
&                    & r\(\leq\)1 & 0.335 & 13.846 $^{**}$ & 13.846 $^{**}$ & \vspace{4pt}\\
cf\_sum, vc\_angel\_seed & 4 & r=0 & 0.554 & 32.815 $^{***}$ & 28.259 $^{***}$ & One coint. vector. \\
&                            & r\(\leq\)1 & 0.122 & 4.556 & 4.556 & \vspace{4pt}\\
cf\_sum, vc\_early\_growth & 5 & r=0 & 0.480 & 36.699 $^{***}$ & 22.207 $^{**}$ & At least one coint. vector. \\
&                              & r\(\leq\)1 & 0.347 & 14.492 $^{**}$ & 14.492  $^{**}$ & \vspace{4pt}\\
\textit{Investment Categories} & \vspace{4pt} \\
cf\_hardware, vc\_hardware & 4 & r=0 & 0.411 & 31.336 $^{***}$ & 18.537 $^{*}$ & At least one coint. vector. \\
&                              & r\(\leq\)1 & 0.306 & 12.799 $^{**}$ & 12.799 $^{**}$ & \vspace{4pt}\\

cf\_media, vc\_media & 5 & r=0 & 0.373 & 23.453 $^{*}$ & 15.885 & No coint. vector. \\
&                        & r\(\leq\)1 & 0.200 & 7.569 & 7.569 & \vspace{4pt}\\

cf\_fashion, vc\_fashion & 5 & r=0 & 0.181 & 13.422 & 6.802 & No coint. vector. \\
&                            & r\(\leq\)1 & 0.177 & 6.620 & 6.620 & \\

\bottomrule
\end{tabular*}
\begin{tablenotes}
\item\footnotesize{$^{***}\ p<0.01$, $^{**}\ p<0.05$, $^{*}\ p<0.1$ significance level. \(\chi ^{2}\) and \(p\)-value are MWald test results. Tested with constant and trend at \(maxlag = 6\). 
The respective null hypothesis is denoted as \(r\) = 0 (“None”) and \(r\) \(\leq\) 1 (“At least one”) cointegration vector(s). The respective null hypothesis is denoted by \(r = 0\) and \(r \leq 1\), where  \(r = 0\) denotes at least one cointegration vector. Optimal lags are based on VAR estimations (AIC, BIC/SIC, FPE), \citet{ljung_box_1978} portmanteau test (asymptotic) and inverse roots of AR characteristics of polynomials for dynamic stability. Trace and Max eigenvalue are the Johansen test statistics. Critical values from \citet{osterwald_lenum_1992}. Critical values for \(r = 0\) at 5\% are 25.32 (Trace) and 18.96 (Max eigenvalue) and 12.25 (Trace, Max eigenvalue) for \(r = 1\).}
\end{tablenotes}
\end{table*}

\subsection{Crowding Out}

We have extensively elaborated on the positive link between crowdfunding and VC investments. Yet, entrepreneurs could also simply substitute formal capital with informal capital through the crowd. This should be relevant especially for smaller and earlier investments. 

In the main analysis, we separate the investments following the CrunchBase definition of angel and seed investments (which we subsequently combine into one subsegment which is sought to include the smaller and earlier investments) and early stage and growth stage investments. Granger causality test revealed a unidirectional causality from crowdfunding to VC investments for angel and seed investments. 

We additionally investigate whether crowding-out effect exists when projects involve smaller amounts of financing using an alternative segmentation based on investment volume. To this aim, we focus on ‘small’ VC investments below \$500,000. When investigating the relation between (all) crowdfunding investments and ‘small’ VC investments, we again do not find any evidence that crowdfunding would crowd out VC investments. We do not detect any significant correlation between the time series. More interestingly, we again document Granger causality from crowdfunding to ‘small’ VC investments (\(\chi ^{2}\)=11.8, \(p\)=0.038).

\section{Discussion}

Our understanding of individual factors enabling and hindering entrepreneurship is steadily increasing. In their interviews, \citet{blanchflower_oswald_1998} provide evidence that inadequate financing is often cited as one of the primary reasons for not becoming self-employed. Over the past years, crowdfunding platforms, such as Kickstarter, have offered financing to several groundbreaking technologies. Pebble, Oculus VR, and Trace have not only garnered media attention, but also attracted seven-digit funding. As \citet{boudreau_lakhani_2013} note, “crowds expand the capabilities of companies; they should be viewed as another tool for organizational problem solving.” Ultimately, successful crowdfunding implies market creation for products whose needs have not been satisfied yet. Essentially, crowdfunding pledges reflect the “wisdom of the crowd” in “screening new venture offerings and voting with their individual investment pledges for the best ones” \citep{bruton_khavul_siegel_wright_2015}. As such, the crowd can also enable VC investors to "nowcast" \citep{choi_varian_2012} new products and markets regarding their proof of concept and potential. Thus, crowdfunding is not only about collecting money from the crowd, but also about crowdsourcing information on needs and demands in the market, as well as, in consequence, new market creation \citep{kim_mauborgne_2014}.

Our results suggest that the crowdfunding volume and the VC investment volume and are strongly correlated, and we document a strong cointegration between the two time series. Moreover, we conclude that an increase in crowdfunding investments Granger causes subsequent VC investments. We corroborate our findings using two VC subsegments: the seed/angel and early/growth stage investments. Here we find a similar and robust pattern: crowdfunding Granger causes VC investments in both segments. Our findings are consistent with the view that successful crowdfunding campaigns help new technologies get attention from other investors. The visible and often large-scale involvement of the online brand community may present a strong signal of a technology viability and market validity, which, in turn, may influence the perception of other investors that have not invested yet. Our study supports the view that crowd is not ‘mad’ but ‘wise’ \citep{mollick_nanda_2015}).

This process may validate the authenticity of a firm and its technology, and may reduce uncertainty as to the feasibility and viability of new technologies on the part of external (later stage) financiers, such as VC investors.  Our findings are also in line with the picture of crowdfunding as a faster and more flexible investment vehicle than VC investments. The exploratory phase may be too risky for these investors and previous works suggest that entrepreneurs usually rely on other sources of capital \citep[e.g.][]{robb_robinson_2014}. Crowdfunding may rapidly adapt to changing economic conditions and to common wide shocks because the fundraising and the decision processes are less complex compared to VC funding \citep{kaplan_stroemberg_2004,mollick_2013}.
The Kickstarter platform has the appealing feature that it combines several sectors (which we summarized into three main investment categories) that are characterized by varying degrees of tangibility and may differ in the longevity of trends involved. Here, interestingly, our results suggest that the Granger causality and cointegration are only present in the hardware category. VC investors tend to invest in the hardware category when it received crowdfunding previously. In areas, such as gaming, internet-related technologies, wearable computing, and three-dimensional printing the crowd appears to be a more reliable predictor \citep{mollick_nanda_2015}. Here, VC investors can move from screening to impacting the organizational development, the hiring of key personnel, and the crafting of human resource policies \citep{hellmann_puri_2002}. In industries were market potentials are higher and processes scalable, VCs may have more margin to maneuver, which is important given the reported inadequacy of commercialization and team building approaches in crowdfunding \citep{mollick_kuppuswamy_2014}.

In the media category, we find a significant correlation, but we do not find evidence for Granger causality and cointegration. We do not find any statistically significant relationship in the fashion category. While the crowd might provide an accurate assessment of trends in industries that involve hardware and technology, it seems to be less likely to provide long-lasting information on industries prone to short-lived trends and fads, such as fashion and media. Investment in these segments may less scalable than hardware and technology investments. As such, there might be limited ways in which the VC investors can help in professionalizing the firm. 
We also test for an alternative explanation reasoning that if the crowd democratizes access to capital it may adversely affect the investment opportunities for VC investors form smaller and earlier investments, but we do not find any evidence that would support the crowding-out argument. In sum, there is no downturn in VC investments when crowd contributions rise. Rather, the crowd and the VC investors seen to be synergistic and complementary members of the same ecosystem.

\section{Conclusion}

Although crowdfunding may be an important predictor of VC investment activity, we know surprisingly little about how the opportunity for entrepreneurial experimentation through crowdfunding affects the demand for VC investments, and the amount that VC investors eventually provide. If the quality and number of entrepreneurs vary across time, then so does the opportunity for VC investors to fund investments. We draw a picture of a nuanced financing eco-system in which entrepreneurial cycles and investment cycles are intertwined. 
Our results suggest that experimentation supported by the crowd can be an important channel through which VC investors obtain access to new technologies and trends. VC-funded startups might jump on a bandwagon of new trends discovered on crowdfunding platforms and commercialize these ideas with more financial resources that make it possible to scale faster. Where the crowd provides signals through aggregated investments, the VC market responds with a lag of four to eight months with proportionally higher investments.\\

\textbf{Acknowledgements}
We like to thank Oleksandr Zastupailo for his support in the data collection and also wish to express a sincere thank you to César A. Hidalgo, Cristian Figueroa Jara, Yves-Alexandre de Montjoye and Dominik Hartmann who provided valuable feedback on earlier versions of this manuscript.

\printbibliography

\appendix
\setcounter{table}{0}  

\renewcommand{\thetable}{A.\arabic{table}}

\begin{table*}[h]
\caption{Kickstarter category matching for CF investment categories.}
\label{tab:appendix1a}
\resizebox{\textwidth}{!}{%
\begin{tabular*}{1.25\textwidth}{lllr}
\toprule
{Kickstarter category} & {Merged category} & {Kickstarter category} & {Merged category}\\
\midrule
fashion                      & Fashion, wellness and personal care & film \& video/animation      & Media, arts and entertainment \\
fashion/accessories          & Fashion, wellness and personal care & film \& video/comedy         & Media, arts and entertainment \\
fashion/apparel              & Fashion, wellness and personal care & film \& video/documentary    & Media, arts and entertainment \\
fashion/childrenswear        & Fashion, wellness and personal care & film \& video/drama          & Media, arts and entertainment \\
fashion/couture              & Fashion, wellness and personal care & film \& video/experimental   & Media, arts and entertainment \\
fashion/footwear             & Fashion, wellness and personal care & film \& video/family         & Media, arts and entertainment \\
fashion/jewelry              & Fashion, wellness and personal care & film \& video/fantasy        & Media, arts and entertainment \\
fashion/pet fashion          & Fashion, wellness and personal care & film \& video/festivals      & Media, arts and entertainment \\
fashion/ready-to-wear        & Fashion, wellness and personal care & film \& video/horror         & Media, arts and entertainment \\
crafts                       & Hardware and consumer electronics   & games/playing cards          & Media, arts and entertainment \\
crafts/candles               & Hardware and consumer electronics   & games/puzzles                & Media, arts and entertainment \\
crafts/crochet               & Hardware and consumer electronics   & games/tabletop games         & Media, arts and entertainment \\
crafts/diy                   & Hardware and consumer electronics   & games/video games            & Media, arts and entertainment \\
crafts/embroidery            & Hardware and consumer electronics   & journalism                   & Media, arts and entertainment \\
crafts/glass                 & Hardware and consumer electronics   & journalism/audio             & Media, arts and entertainment \\
crafts/knitting              & Hardware and consumer electronics   & journalism/photo             & Media, arts and entertainment \\
crafts/letterpress           & Hardware and consumer electronics   & journalism/print             & Media, arts and entertainment \\
crafts/pottery               & Hardware and consumer electronics   & journalism/video             & Media, arts and entertainment \\
crafts/printing              & Hardware and consumer electronics   & journalism/web               & Media, arts and entertainment \\
crafts/quilts                & Hardware and consumer electronics   & music                        & Media, arts and entertainment \\
crafts/stationery            & Hardware and consumer electronics   & music/blues                  & Media, arts and entertainment \\
crafts/taxidermy             & Hardware and consumer electronics   & music/chiptune               & Media, arts and entertainment \\
crafts/weaving               & Hardware and consumer electronics   & music/classical music        & Media, arts and entertainment \\
crafts/woodworking           & Hardware and consumer electronics   & music/country \& folk        & Media, arts and entertainment \\
design                       & Hardware and consumer electronics   & music/electronic music       & Media, arts and entertainment \\
design/architecture          & Hardware and consumer electronics   & music/faith                  & Media, arts and entertainment \\
design/civic design          & Hardware and consumer electronics   & music/hip-hop                & Media, arts and entertainment \\
design/product design        & Hardware and consumer electronics   & music/indie rock             & Media, arts and entertainment \\
technology                   & Hardware and consumer electronics   & music/jazz                   & Media, arts and entertainment \\
technology/3d printing       & Hardware and consumer electronics   & music/kids                   & Media, arts and entertainment \\
technology/camera equipment  & Hardware and consumer electronics   & music/latin                  & Media, arts and entertainment \\
technology/diy electronics   & Hardware and consumer electronics   & music/metal                  & Media, arts and entertainment \\
technology/fabrication tools & Hardware and consumer electronics   & music/pop                    & Media, arts and entertainment \\
technology/flight            & Hardware and consumer electronics   & music/punk                   & Media, arts and entertainment \\
technology/gadgets           & Hardware and consumer electronics   & music/r\&b                   & Media, arts and entertainment \\
technology/hardware          & Hardware and consumer electronics   & music/rock                   & Media, arts and entertainment \\
technology/makerspaces       & Hardware and consumer electronics   & music/world music            & Media, arts and entertainment \\
technology/robots            & Hardware and consumer electronics   & photography                  & Media, arts and entertainment \\
technology/sound             & Hardware and consumer electronics   & photography/animals          & Media, arts and entertainment \\
technology/space exploration & Hardware and consumer electronics   & photography/fine art         & Media, arts and entertainment \\
technology/wearables         & Hardware and consumer electronics   & photography/nature           & Media, arts and entertainment \\
art                          & Media, arts and entertainment       & photography/people           & Media, arts and entertainment \\
art/ceramics                 & Media, arts and entertainment       & photography/photobooks       & Media, arts and entertainment \\
art/conceptual art           & Media, arts and entertainment       & photography/places           & Media, arts and entertainment \\
art/digital art              & Media, arts and entertainment       & publishing                   & Media, arts and entertainment \\
art/illustration             & Media, arts and entertainment       & publishing/academic          & Media, arts and entertainment \\
art/installations            & Media, arts and entertainment       & publishing/anthologies       & Media, arts and entertainment \\
art/mixed media              & Media, arts and entertainment       & publishing/art books         & Media, arts and entertainment \\
art/painting                 & Media, arts and entertainment       & publishing/calendars         & Media, arts and entertainment \\
art/performance art          & Media, arts and entertainment       & publishing/children's books  & Media, arts and entertainment \\
art/public art               & Media, arts and entertainment       & publishing/fiction           & Media, arts and entertainment \\
art/sculpture                & Media, arts and entertainment       & publishing/literary journals & Media, arts and entertainment \\
art/textiles                 & Media, arts and entertainment       & publishing/nonfiction        & Media, arts and entertainment \\
art/video art                & Media, arts and entertainment       & publishing/periodicals       & Media, arts and entertainment \\
\bottomrule
\end{tabular*}
}
\end{table*}

\pagebreak

\begin{table*}[h]
\caption{Kickstarter category matching for CF investment categories (continued).}
\label{tab:appendix1b}
\resizebox{\textwidth}{!}{%
\begin{tabular*}{1.25\textwidth}{lllr}
\toprule
{Kickstarter category} & {Merged category} & {Kickstarter category} & {Merged category}\\
\midrule
comics                       & Media, arts and entertainment       & publishing/poetry            & Media, arts and entertainment \\
comics/anthologies           & Media, arts and entertainment       & publishing/radio \& podcasts & Media, arts and entertainment \\
comics/comic books           & Media, arts and entertainment       & publishing/translations      & Media, arts and entertainment \\
comics/events                & Media, arts and entertainment       & publishing/young adult       & Media, arts and entertainment \\
comics/graphic novels        & Media, arts and entertainment       & publishing/zines             & Media, arts and entertainment \\
comics/webcomics             & Media, arts and entertainment       & technology/apps              & Media, arts and entertainment \\
dance                        & Media, arts and entertainment       & technology/software          & Media, arts and entertainment \\
dance/performances           & Media, arts and entertainment       & technology/web               & Media, arts and entertainment \\
dance/residencies            & Media, arts and entertainment       & theater                      & Media, arts and entertainment \\
dance/spaces                 & Media, arts and entertainment       & theater/experimental         & Media, arts and entertainment \\
dance/workshops              & Media, arts and entertainment       & theater/festivals            & Media, arts and entertainment \\
design/graphic design        & Media, arts and entertainment       & theater/immersive            & Media, arts and entertainment \\
design/interactive design    & Media, arts and entertainment       & theater/musical              & Media, arts and entertainment \\
design/typography            & Media, arts and entertainment       & theater/plays                & Media, arts and entertainment \\
film \& video                & Media, arts and entertainment       & theater/spaces               & Media, arts and entertainment \\
film \& video/action         & Media, arts and entertainment       &                              &                              \\
\bottomrule
\end{tabular*}
}
\end{table*}

\pagebreak

\begin{table*}[h]
\caption{Crunchbase category matching for VC investment categories.}
\label{tab:appendix2a}
\resizebox{\textwidth}{!}{%
\begin{tabular*}{1.25\textwidth}{lllr}
\toprule
{CrunchBase category} & {Merged category} & {CrunchBase category} & {Merged category}\\
\midrule

Babies                     & Fashion, wellness and personal care & Creative                    & Media, arts and entertainment \\
Beauty                     & Fashion, wellness and personal care & Curated Web                 & Media, arts and entertainment \\
Cosmetics                  & Fashion, wellness and personal care & Design                      & Media, arts and entertainment \\
Eyewear                    & Fashion, wellness and personal care & Digital Media               & Media, arts and entertainment \\
Fashion                    & Fashion, wellness and personal care & Entertainment               & Media, arts and entertainment \\
Fitness                    & Fashion, wellness and personal care & Entertainment Industry      & Media, arts and entertainment \\
Health and Wellness        & Fashion, wellness and personal care & Events                      & Media, arts and entertainment \\
Jewelry                    & Fashion, wellness and personal care & Facebook Applications       & Media, arts and entertainment \\
Kids                       & Fashion, wellness and personal care & Fantasy Sports              & Media, arts and entertainment \\
Lifestyle                  & Fashion, wellness and personal care & File Sharing                & Media, arts and entertainment \\
Lifestyle Products         & Fashion, wellness and personal care & Film                        & Media, arts and entertainment \\
Mobile Health              & Fashion, wellness and personal care & Gambling                    & Media, arts and entertainment \\
Mothers                    & Fashion, wellness and personal care & Game                        & Media, arts and entertainment \\
Personal Health            & Fashion, wellness and personal care & Game Mechanics              & Media, arts and entertainment \\
Sporting Goods             & Fashion, wellness and personal care & Games                       & Media, arts and entertainment \\
Women                      & Fashion, wellness and personal care & Gamification                & Media, arts and entertainment \\
3D                         & Hardware and consumer electronics   & Maps                        & Media, arts and entertainment \\
3D Printing                & Hardware and consumer electronics   & Media                       & Media, arts and entertainment \\
3D Technology              & Hardware and consumer electronics   & Messaging                   & Media, arts and entertainment \\
Augmented Reality          & Hardware and consumer electronics   & MicroBlogging               & Media, arts and entertainment \\
Auto                       & Hardware and consumer electronics   & Mobile Games                & Media, arts and entertainment \\
Automotive                 & Hardware and consumer electronics   & Mobile Video                & Media, arts and entertainment \\
Batteries                  & Hardware and consumer electronics   & Mobility                    & Media, arts and entertainment \\
Bicycles                   & Hardware and consumer electronics   & Music                       & Media, arts and entertainment \\
Bitcoin                    & Hardware and consumer electronics   & Music Education             & Media, arts and entertainment \\
Cars                       & Hardware and consumer electronics   & Music Services              & Media, arts and entertainment \\
Communications Hardware    & Hardware and consumer electronics   & Music Venues                & Media, arts and entertainment \\
Computers                  & Hardware and consumer electronics   & Musicians                   & Media, arts and entertainment \\
Consumer Electronics       & Hardware and consumer electronics   & News                        & Media, arts and entertainment \\
Consumer Goods             & Hardware and consumer electronics   & Online Dating               & Media, arts and entertainment \\
Displays                   & Hardware and consumer electronics   & Online Gaming               & Media, arts and entertainment \\
Drones                     & Hardware and consumer electronics   & Peer-to-Peer                & Media, arts and entertainment \\
Electric Vehicles          & Hardware and consumer electronics   & Photo Sharing               & Media, arts and entertainment \\
Electronics                & Hardware and consumer electronics   & Photography                 & Media, arts and entertainment \\
Energy                     & Hardware and consumer electronics   & Portals                     & Media, arts and entertainment \\
Hardware                   & Hardware and consumer electronics   & Privacy                     & Media, arts and entertainment \\
Hardware + Software        & Hardware and consumer electronics   & Publishing                  & Media, arts and entertainment \\
Human Computer Interaction & Hardware and consumer electronics   & Real Time                   & Media, arts and entertainment \\
Internet of Things         & Hardware and consumer electronics   & Reviews and Recommendat. & Media, arts and entertainment \\
Lasers                     & Hardware and consumer electronics   & Semantic Web                & Media, arts and entertainment \\
Lighting             & Hardware and consumer electronics & Social Bookmarking   & Media, arts and entertainment \\
Mac                  & Hardware and consumer electronics & Social Games         & Media, arts and entertainment \\
Mechanical Solutions & Hardware and consumer electronics & Social Media         & Media, arts and entertainment \\
Medical              & Hardware and consumer electronics & Social Network Media & Media, arts and entertainment \\
Medical Devices      & Hardware and consumer electronics & Social Television    & Media, arts and entertainment \\
MMO Games            & Hardware and consumer electronics & Sports               & Media, arts and entertainment \\
Mobile Devices       & Hardware and consumer electronics & Tablets              & Media, arts and entertainment \\
New Technologies     & Hardware and consumer electronics & Television           & Media, arts and entertainment \\
Robotics             & Hardware and consumer electronics & Theatre              & Media, arts and entertainment \\
Sensors              & Hardware and consumer electronics & Ticketing            & Media, arts and entertainment \\
Solar                & Hardware and consumer electronics & Toys                 & Media, arts and entertainment \\
Telephony            & Hardware and consumer electronics & TV Production        & Media, arts and entertainment \\
Wireless             & Hardware and consumer electronics & Twitter Applications & Media, arts and entertainment \\
Android              & Media, arts and entertainment     & Video                & Media, arts and entertainment \\
\bottomrule
\end{tabular*}
}
\end{table*}

\pagebreak

\begin{table*}[h]
\caption{Crunchbase category matching for VC investment categories (continued).}
\label{tab:appendix2b}
\resizebox{\textwidth}{!}{%
\begin{tabular*}{1.25\textwidth}{lllr}
\toprule
{CrunchBase category} & {Merged category} & {CrunchBase category} & {Merged category}\\
\midrule
Apps                 & Media, arts and entertainment     & Video Chat           & Media, arts and entertainment \\
Art                  & Media, arts and entertainment     & Video Editing        & Media, arts and entertainment \\
Audio                & Media, arts and entertainment     & Video Games          & Media, arts and entertainment \\
Blogging Platforms   & Media, arts and entertainment     & Video on Demand      & Media, arts and entertainment \\
Broadcasting         & Media, arts and entertainment     & Video Streaming      & Media, arts and entertainment \\
Celebrity            & Media, arts and entertainment     & Virtual Goods        & Media, arts and entertainment \\
Chat                 & Media, arts and entertainment     & Virtual Worlds       & Media, arts and entertainment \\
Communities          & Media, arts and entertainment     & Virtualization       & Media, arts and entertainment \\
Concerts             & Media, arts and entertainment     & Visualization        & Media, arts and entertainment \\
Content              & Media, arts and entertainment     & Windows Phone 7      & Media, arts and entertainment \\
Content Creators     & Media, arts and entertainment     & Writers              & Media, arts and entertainment \\
\bottomrule
\end{tabular*}
}
\end{table*}

\end{document}